\documentclass[prd,preprint]{revtex4}
\usepackage{amsmath}
\usepackage{latexsym}
\usepackage{amsfonts}
\begin{document}
\title{Quintessence Model and Observational Constraints}
\author{Yungui Gong} \email{ygong@ece.ogi.edu}
\affiliation{Department of Electrical and Computer Engineering,
Oregon Graduate Institute, Beaverton, OR 97006}
\begin{abstract}
The recent observations of type Ia supernovae strongly support
that the universe is accelerating now and decelerated in the
recent past. By assuming a general relation between the
quintessence potential and the quintessence kinetic energy, a
general relation is found between the quintessence energy density
and the scale factor. The potential includes both the hyperbolic
and the double exponential potentials. A detailed analysis of the
transition from the deceleration phase to the acceleration phase
is then performed. We show that the current constraints on the
transition time, the equation of state and the energy density of
the quintessence field are satisfied in the model.
\end{abstract}
\pacs{98.80.Cq, 98.80.Es}
\maketitle

\parindent=4ex

The recent observations of Type Ia supernovae strongly support
that the expansion of the Universe is accelerating
\cite{sp97}\cite{agr98}. The observation of type Ia supernova SN
1997ff at $z\sim 1.7$ also provides the evidence of a decelerated
universe in the recent past \cite{agr}. Turner and Riess showed
that the supernova data favored recent acceleration ($z<0.5$) and
past deceleration ($z>0.5$) \cite{mstagr}. On the other hand, the
measurement of the acoustic peaks in the angular power spectrum of
the cosmic microwave background supports a flat universe as
predicted by the inflationary models \cite{pdb00}\cite{sh00}. One
immediate conclusion is that the universe is dominated by a
form of matter with negative pressure now, this form of matter is
widely referred as dark energy. One simple candidate of the dark
energy is the cosmological constant. However, there are some
problems although cold dark matter cosmological constant models
are consistent with the current observations. Why does the vacuum
energy begin to dominate presently? This problem is referred
as the coincident problem. Why is the cosmological constant so
small and not zero? The quintessence models avoid these problems
\cite{cds98}-\cite{mcbobncs}. The basic idea of the quintessence
model bases upon a scalar field $Q$ that slowly evolves down its potential
$V(Q)$. The slowly evolved scalar field takes the role of
a dynamical cosmological constant. The energy density of the scalar
field must remain very small compared with radiation and matter at
early epoches and evolves in a way that it started to dominate the
universe at very recent past. This requires some constraints on
the initial conditions and fine tuning of the potential.  A
tracker field was used to solve the initial condition and fine
tuning problems. The missing energy was comparable to the
radiation energy at the very early time, tracked the background
energy density for most of the history of the universe, and then
grew to dominate the energy density recently so that the Universe
is accelerating now. Although these models successfully explain
the dark energy of the universe, they fail to discuss matching
the quintessence field dominating the accelerating universe with the matter
dominating the decelerating universe. Sen and Sethi addressed the
smoothly matching issue by assuming a particular form of the scale
factor \cite{aasss}.  In this paper, we assumed a type of
potential that has a general relation with the kinetic term of the
quintessence field and then obtained  an exact relation between
the energy density of the quintessence field and the scale factor.
This type of solution includes the constant pressure solution
corresponding to the double exponential potential given in
\cite{aasss} and the constant equation of state solution
corresponding to hyperbolic potential discussed in
\cite{johri}\cite{lautm}. The constraints on the current values of
$\omega_{Q}$ and $\Omega_{Q}$ given in \cite{rbsham} were then
used to show that the model passed those constraints. Those
constraints disfavored some quintessence models, like the inverse
power law potential \cite{pscejc}\cite{wetterich}\cite{admcs}.

For a spatially flat, isotropic and homogeneous universe with both
an ordinary pressureless dust matter and a minimally coupled
scalar field $Q$ sources, the Friedmann equations are
\begin{gather}
\label{cos1} H^2=\left({\dot{a}\over a}\right)^2={8\pi
G\over 3}(\rho_m+\rho_Q),\\
\label{cos2}
{\ddot{a}\over a}=-{4\pi G\over
3}(\rho_m+\rho_Q+3p_Q),\\
\label{cos3} \ddot{Q}+3H\dot{Q}+V'(Q)=0,
\end{gather}
where dot means derivative with respect to time,
$\rho_m=\rho_{m0}(a_0/a)^3$ is the matter energy density, a
subscript 0 means the value of the variable at present time,
$\rho_Q=\dot{Q}^2/2+V(Q)$, $p_Q=\dot{Q}^2/2-V(Q)$,
$V'(Q)=dV(Q)/dQ$ and $V(Q)$ is the potential of the quintessence
field. To proceed, we assume the following general relationship
\begin{equation}
\label{assm1}
V(Q)=\beta \dot{Q}^2+C,
\end{equation}
instead of assuming a particular potential for the quintessence
field or a particular form of the scale factor, where $\beta$ and
$C$ are constants. The above general potential includes the
hyperbolic potential and the double exponential potential. From
Eq. (\ref{assm1}), we get
\begin{equation}
\label{rel1}
V'(Q)=2\beta\ddot{Q}.
\end{equation}
Substitute Eq. (\ref{rel1})  to Eq. (\ref{cos3}), we get
\begin{equation}
\label{rel2} \dot{Q}a^{3/(2\beta+1)}=C_2,
\end{equation}
where $C_2$ is an integration constant. Therefore, the energy
density and pressure of the quintessence field become
\begin{gather}
\label{eng} \rho_Q={(1/2+\beta)C^2_2\over a^{6/(2\beta+1)}}+C,\\
\label{qup} p_Q={(1/2-\beta)C^2_2\over a^{6/(2\beta+1)}}-C.
\end{gather}
The equation of state of the quintessence field is
$$\omega_Q={(1/2-\beta)C^2_2-Ca^{6/(2\beta+1)}\over
(1/2+\beta)C^2_2+Ca^{6/(2\beta+1)}}.$$ In terms of $\rho_{Q0}$ and
$\omega_{Q0}$, we have
\begin{gather}
C_2^2=(1+\omega_{Q0})\rho_{Q0}a^{6/(2\beta+1)},\\
C=\left[{1\over 2}-\beta-\left({1\over
2}+\beta\right)\omega_{Q0}\right]\rho_{Q0},\\
\label{eng1}
\rho_Q=(1/2+\beta)(1+\omega_{Q0})\rho_{Q0}\left({a_0\over
a}\right)^{6/(2\beta+1)} +\left[{1\over 2}-\beta-\left({1\over
2}+\beta\right)\omega_{Q0}\right]\rho_{Q0},\\
\label{qup1}
p_Q=(1/2-\beta)(1+\omega_{Q0})\rho_{Q0}\left({a_0\over
a}\right)^{6/(2\beta+1)} -\left[{1\over 2}-\beta-\left({1\over
2}+\beta\right)\omega_{Q0}\right]\rho_{Q0},\\
\label{hub} H^2={8\pi G\over 3}\left\{\rho_{m0}\left(a_0\over
a\right)^3+(1/2+\beta)(1+\omega_{Q0})\rho_{Q0}\left({a_0\over
a}\right)^{6/(2\beta+1)} +C\right\}.
\end{gather}
The transition from deceleration to acceleration happens when the
deceleration parameter $q=-\ddot{a}H^2/a=0$. From Eqs.
(\ref{cos2}), (\ref{eng1}) and (\ref{qup1}), in terms of the
redshift parameter $1+z=a_0/a$, we have
\begin{equation}
\label{rel3}
(1+z_{q=0})^3+2(1-\beta)(1+\omega_{Q0}){\rho_{Q0}\over
\rho_{m0}}(1+z_{q=0})^{6/(2\beta+1)}-[1-2\beta-(1+2\beta)\omega_{Q0}]{\rho_{Q0}\over
\rho_{m0}}=0
\end{equation}
This equation gives a relationship between $\omega_{Q0}$ and
$\Omega_{Q0}$. Now let us look at some special cases.

Case I: $C=0$. This is the case that the equation of state of the
scalar field is a constant, $\omega_Q=(1/2-\beta)/(1/2+\beta)$. The potential is \cite{johri}
$$V(Q)=A[\sinh k(Q/\alpha+B)]^{-\alpha},$$
where $\alpha=2/(\beta-1/2)$, $k^2=48\pi G/(2\beta+1)$,
$A^{\beta-1/2}=(1/2+\beta)C_2^{2\beta+1}\beta^{\beta-1/2}/(\rho_{m0}a^3_0)$
and $B$ is an arbitrary integration constant. The energy density
of the quintessence field evolves as
$$\rho_Q=\rho_{Q0}\left({a_0\over
a}\right)^{3(1+\omega_{Q0})}.$$ Eq. (\ref{rel3}) becomes
$$(1+z_{q=0})^3+{\rho_{Q0}\over
\rho_{m0}}(1+3\omega_{Q0})(1+z_{q=0})^{3(1+\omega_{Q0})}=0.$$ Use
the observation results $\rho_{Q0}/\rho_{m0}=7/3$ and
$z_{q=0}=0.5$, we get $\omega_{Q0}=-0.65$. However, if we take
$z_{q=0}=0.666$, then $\omega_{Q0}=-0.9$. If we take $z_{q=0}=0.5$
and $\rho_{Q0}/\rho_{m0}=0.64/0.36=1.8$, then $\omega_{Q0}=-0.88$.
This shows that the transition time $z_{q=0}$ from deceleration to
acceleration, $\omega_{Q0}$ and $\Omega_{Q0}$ are very sensitive
to each other. The recent constraint is $\omega_{Q0}<-0.85$.

Case II: $\beta=1/2$. This is the case discussed in \cite{aasss} .
This case gives a constant pressure $p_Q=\omega_{Q0}\rho_{Q0}$ for
the scalar field and the potential is the double exponential
potential. The solutions to Eqs. (\ref{eng1}) and (\ref{hub}) are
\begin{gather}
\label{eng2} \rho_Q=(1+\omega_{Q0})\rho_{Q0}\left({a_0\over
a}\right)^3 -\omega_{Q0}\rho_{Q0}=
  \begin{cases}
    (1+\omega_{Q0})\rho_{Q0}\left({a_0\over
a}\right)^3 & a_0/a\gg 1, \\
    -\omega_{Q0}\rho_{Q0} & a_0/a \ll 1.
  \end{cases}
,\\
\rho_{Q0}+\rho_{m0}={1\over 6\pi Gt^2_0}[\coth(1)]^2,\\
a(t)={a_0\over [\sinh(1)]^{2/3}}[\sinh(t/t_0)]^{2/3}.
\end{gather}
During matter dominated epoch, the quintessence field should be
sub-dominated and this gives the constraint
$(1+\omega_{Q0})\rho_{Q0}< \rho_{m0}$. Now Eq. (\ref{rel3})
becomes
$$(1+z_{q=0})^3+(1+\omega_{Q0}){\rho_{Q0}\over
\rho_{m0}}(1+z_{q=0})^3+2\omega_{Q0}{\rho_{Q0}\over
\rho_{m0}}=0.$$ Use $z_{q=0}=0.5$ and $\rho_{Q0}/\rho_{m0}=7/3$,
we get $\omega_{Q0}=-0.897$ and
$(1+\omega_{Q0})\rho_{Q0}/\rho_{m0}=0.24$. If we use $z_{q=0}\sim
0.67$ or $\rho_{Q0}/\rho_{m0}\sim 0.63/0.37=1.7$, then
$\omega_{Q0}\sim -1$. Again the transition time $z_{q=0}$ from
deceleration to acceleration, $\omega_{Q0}$ and $\Omega_{Q0}$ are
very sensitive to each other.

Case III: $\beta=1$. The energy density of the scalar field
evolves as
$$\rho_{Q}={3\over 2}(1+\omega_{Q0})\rho_{Q0}\left({a_0\over
a}\right)^2-{1\over 2}(1+3\omega_{Q0})\rho_{Q0}.$$ The evolution
of the Universe is the same as that of $k=-1$ with a cosmological
constant. The constraint equation (\ref{rel3}) becomes
$$(1+z_{q=0})^3+(1+3\omega_{Q0}){\rho_{Q0}\over
\rho_{m0}}=0.$$ Take $z_{q=0}=0.5$ and $\rho_{Q0}/\rho_{m0}=7/3$,
we get $\omega_{Q0}=-0.82$.

It is  also interesting to note that for a pure cosmological
constant ($C_2=0$ in our model), we have the relation
$$(1+z_{q=0})^3={2\rho_{Q0}\over \rho_{m0}}.$$

With the assumption of Eq. (\ref{assm1}), we are able to get the
energy density Eq. (\ref{eng1}) of the quintessence field. The
energy density of the quintessence field has the property that it
is sub-dominated at higher redshift and becomes dominated at
present time. This solution makes it possible to study the
detailed transition from deceleration phase to acceleration phase.
The model also gives the relation among $z_{q=0}$, $\omega_{Q0}$
and $\Omega_{Q0}$. The constraints $\omega_{Q0}<-0.85$,
$\Omega_{Q0}=0.65\pm 0.15$ and $z_{q=0}\sim 0.5$ are satisfied in
the model. If $\omega_{Q0}\rightarrow -1$, then $\Omega_{Q0}$
tends to take lower value or $z_{q=0}$ tends to take higher value.
The model can also easily pass the constraint
$\Omega_Q($Mev$)<0.045$.

\begin{acknowledgments}
The author would like to thank Professors John Barrow and Alexei
Starobinsky for bringing their works into the author's attention.
\end{acknowledgments}

\end{document}